\documentstyle[12pt,psfig]{article}

\makeatletter
\@addtoreset{equation}{section}

\def\baselinestretch{1.2}
\def\href#1#2{#2}  

\newcommand{\norm}[1]{\raise.3ex\hbox{:} #1 \raise.3ex\hbox{:}\,}

\def\tr{{\rm Tr}}
\def\GCD{{\rm GCD}}
\def\im{{\rm Im}}
\def\re{{\rm Re}}

\newcommand{\beq}{\begin{equation}}
\newcommand{\eeq}{\end{equation}}
\newcommand{\al}{\alpha^{'}}

\newcommand{\Th}{\Theta}

\begin{document}

\begin{titlepage}

\begin{flushright}
NSF-ITP-99-133\\
hep-th/9911057
\end{flushright}
\vfil\vfil

\begin{center}

{\Large {\bf On the Hierarchy Between  Non-Commutative
 and  Ordinary Supersymmetric Yang-Mills}}

\vfil

\vspace{10mm}

Akikazu Hashimoto\\

\vspace{10mm}

Institute for Theoretical Physics\\ University of California,
Santa Barbara, CA 93106\\
aki@itp.ucsb.edu\\

\vspace{10mm}

N. Itzhaki\\

\vspace{10mm}

Department of Physics\\ University of California,
Santa Barbara, CA 93106\\
sunny@physics.ucsb.edu\\

\vfil
\end{center}

\begin{abstract}
\noindent Non-commutative supersymmetric Yang-Mills with rational
$\Theta$ is dual to an ordinary supersymmetric Yang-Mills with a 't
Hooft flux. It is believed that the simplest description is via the
ordinary supersymmetric Yang-Mills. We claim, however, that
the two descriptions form a hierarchy. The SYM description is the
proper description in the ultra violet while the non-commutative
description takes over in the infra red.
\end{abstract}

\vfil\vfil\vfil

\end{titlepage}
\renewcommand{\baselinestretch}{1.05}  

\section{Introduction}

It is well known that a non-commutative supersymmetric Yang-Mills
theory (NCSYM) with non-commutativity scale $\Delta$ on a torus of
size $\Sigma$ is equivalent to an ordinary supersymmetric Yang-Mills
theory (OSYM) with a magnetic flux if the dimensionless
non-commutativity parameter $\Theta = \Delta^2/\Sigma^2$ is a rational
number.  The equivalence between the NCSYM and the OSYM can be
understood naturally as arising from the T-duality of the underlying
string theory \cite{Connes:1998cr,Douglas:1998fm}.  The structure of
T-duality survives the decoupling limit and can be identified as the
Morita equivalence of the NCSYM. The precise relation between
T-duality and Morita equivalence was studied extensively
\cite{Ho:1998hq,Bigatti:1998vf,Landi:1998ii,Morariu:1998qm,Hofman:1998pd,Hofman:1998iy,Hofman:1998iv,Schwarz:1998qj,Konechny:1998wv,Konechny:1999rz,Konechny:1999tu,Pioline:1999xg,Seiberg:1999vs}.

Whenever a theory admits dual descriptions, there is always a unique
canonical description that is in some sense natural. For example, in
the case of duality relation which exchanges strong and weak
couplings, it is more natural to describe the physics in terms of the
weakly coupled theory. This is not a useful criterion for our purposes
since Morita equivalence is not a strong-weak duality but rather a
local-nonlocal duality.  That is, it takes a local excitation of one
description to a non-local excitation of the dual description.  This
suggests that the natural description is the one which is most local.

In light of this criterion, one might be inclined to conclude that the
OSYM description is the ultimately local and therefore the most
natural description of this theory.  In that case, one can forget all
together about the NCSYM with rational $\Th$.  However, in the
presence of the magnetic flux, there are very light excitations, the
electric fluxes, which are highly non-local.  The locality of the OSYM
description breaks down when these non-local excitations start to
dominate the dynamics, and this happens precisely in the infra red at
the energies of the order of the volume of the torus.

The primary goal of this paper is to demonstrate that the dynamics of
these light non-local degrees of freedom of the OSYM admits a natural
local description in terms of the NCSYM. In other words, the NCSYM
takes over as the natural local description in the infra red when the
locality of the OSYM breaks down due to finite size and the magnetic
flux. This is the sense in which the NCSYM description is physically
significant and should therefore not be dismissed even if the theory
is equivalent to some OSYM by duality. NCSYM and OSYM are the
components in the hierarchy of the phases of the same theory.

The paper is organized as follows.  In section 2 we describe the
OSYM-NCSYM hierarchy using weakly coupled field theory arguments.  We
also show how similar structures emerge from the supergravity
description of these theories in the limit of large 't Hooft coupling.
Concluding remarks and applications are presented in section 3.

\section{NCSYM as the low energy limit of OSYM}

In this section, we will show how different Morita equivalent theories
form a hierarchy of phases, and that for any given energies, there is
a unique simple/natural description.

We start with 
the simplest case of  NCSYM with $\Th =1/s$ for some
integer $s$. There are two natural Morita equivalent descriptions
 of this
theory: the NCSYM description in the infra red and the OSYM in the
ultra violet. In section 2.1 we show this for a weakly coupled theory
while in section 2.2 we obtain the same result for the strongly
coupled theory using the supergravity description.

The general rational case $\Th =r/s$ is considered in section 2.3.  It
is shown that as we increase the energy, the non-commutativity scale
of the natural description decreases until we end up with some OSYM
description in the deep ultra violet.  We also comment briefly on the
irrational case.

\subsection{NCSYM with $\Theta = 1/s$}

Consider a $U(N)$ NCSYM with coupling constant $g_{YM}$ and vanishing
magnetic flux ($m=0$) on a torus of size $\Sigma$ with $\Theta = 1/s$
for some integer $s\gg 1$.  This theory is equivalent to an ordinary
SYM with
\beq\label{d2}
\tilde N = s N,\qquad\tilde \Sigma =\frac{\Sigma}{s},\qquad\;
\tilde g_{YM}^2
=\frac{g_{YM}^2}{s},\qquad \tilde m =N.
\eeq
The relevant properties of the T-duality (Morita equivalence) used to
derive this relation are summarized in the appendix.
 
It may seem strange that a theory defined on a torus whose volume is
smaller by a factor of $1/s^2 \ll 1$ can be equivalent to a NCSYM on a
much larger volume $\Sigma^2$, since the momentum modes in the
ordinary theory have a gap of the order $1/\tilde \Sigma =s/\Sigma$
while the momentum modes in the non-commutative description have a gap
of the order $1/\Sigma$. However, the fact that the background has a
non-vanishing magnetic flux $\tilde m$ (or a twisted boundary
condition \cite{'tHooft:1981sz}) plays a significant role here.  In a
presence of such a twist, the energies for the excitations
corresponding to the electric fluxes winding around the torus are
quantized in the units of
\beq { \GCD (\tilde m, \tilde N) \over \tilde N} {1 \over \tilde
\Sigma} = {1 \over s \tilde{\Sigma}}.  \eeq 
This was shown in \cite{vanBaal:1984ar} by studying the small
fluctuation of gauge fields in the 't Hooft flux background
\cite{'tHooft:1981sz} and can also be seen from the string theory
considerations \cite{Hashimoto:1997gm}.  This can also be thought of
as a realization of the mechanism of \cite{Maldacena:1996ds}. In this
way, the OSYM in a twisted background and the NCSYM can have the same
physical content.

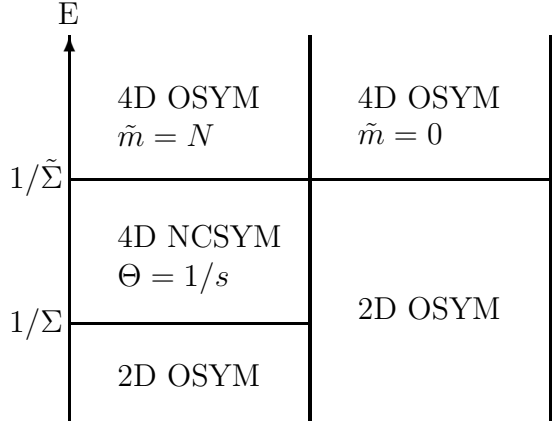
\begin{figure}
\centerline{\setlength{\unitlength}{0.00083300in}%
\begingroup\makeatletter\ifx\SetFigFont\undefined%
\gdef\SetFigFont#1#2#3#4#5{%
  \reset@font\fontsize{#1}{#2pt}%
  \fontfamily{#3}\fontseries{#4}\fontshape{#5}%
  \selectfont}%
\fi\endgroup%
\begin{picture}(3387,2595)(1726,-2773)
\thicklines
\put(2101,-2161){\line( 1, 0){1500}}
\put(2101,-1261){\line( 1, 0){3000}}
\put(3601,-361){\line( 0,-1){2400}}
\put(5101,-361){\line( 0,-1){2400}}
\put(2101,-2761){\vector( 0, 1){2400}}
\put(2401,-2551){\makebox(0,0)[lb]{\smash{\SetFigFont{12}{14.4}{\familydefault}{\mddefault}{\updefault}2D OSYM}}}
\put(2401,-1681){\makebox(0,0)[lb]{\smash{\SetFigFont{12}{14.4}{\familydefault}{\mddefault}{\updefault}4D NCSYM}}}
\put(2401,-1921){\makebox(0,0)[lb]{\smash{\SetFigFont{12}{14.4}{\familydefault}{\mddefault}{\updefault}$\Theta=1/s$}}}
\put(2401,-811){\makebox(0,0)[lb]{\smash{\SetFigFont{12}{14.4}{\familydefault}{\mddefault}{\updefault}4D OSYM}}}
\put(2401,-1051){\makebox(0,0)[lb]{\smash{\SetFigFont{12}{14.4}{\familydefault}{\mddefault}{\updefault}$\tilde m=N$}}}
\put(3901,-811){\makebox(0,0)[lb]{\smash{\SetFigFont{12}{14.4}{\familydefault}{\mddefault}{\updefault}4D OSYM}}}
\put(3901,-1036){\makebox(0,0)[lb]{\smash{\SetFigFont{12}{14.4}{\familydefault}{\mddefault}{\updefault}$\tilde m=0$}}}
\put(3901,-2131){\makebox(0,0)[lb]{\smash{\SetFigFont{12}{14.4}{\familydefault}{\mddefault}{\updefault}2D OSYM}}}
\put(2028,-286){\makebox(0,0)[lb]{\smash{\SetFigFont{12}{14.4}{\familydefault}{\mddefault}{\updefault}E}}}
\put(1726,-1313){\makebox(0,0)[lb]{\smash{\SetFigFont{12}{14.4}{\familydefault}{\mddefault}{\updefault}$1/\tilde{\Sigma}$}}}
\put(1726,-2213){\makebox(0,0)[lb]{\smash{\SetFigFont{12}{14.4}{\familydefault}{\mddefault}{\updefault}$1/\Sigma$}}}
\end{picture}
}
\caption{The phases of SYM on $R^2 \times T^2$: In the absence of a
magnetic flux the theory flows to a two dimensional theory at energies
below $1/\tilde \Sigma$. In the presence of a magnetic flux there are
excitations which are lighter than $1/\tilde \Sigma$. The dynamics of
these excitations is described by the NCSYM in four dimensions. Only
deeper in the infra red the theory becomes two
dimensional.\label{figaa}}
\end{figure}

We will now argue that the NCSYM and the OSYM descriptions provide a
natural description of the dynamics in a non-overlapping range of
energies (see figure \ref{figaa}). Consider first the NCSYM.  The
range of energies where this description is natural is bounded both in
the ultra violet and in the infra red.  The infra red bound is the
usual one which appears also for local theories.  Namely, by lowering
the energies to be of order $1/\Sigma$, we reach energies of the order
of the mass gap of the momentum modes and the effective description is
in terms of the 2D OSYM.

The bound in the ultra violet is more interesting because ordinarily,
the finite size do not affect the ultra violet.  This is not the case
with theories on non-commutative spaces.  By going to higher energies,
one might succeed in making the Compton wavelength $\Delta x_2 = 1/E$
along, say, the 2 direction to be small.  This will cause the size of
the excitations to grow in the 3-direction in keeping with the
non-commutativity $\Delta x_2 \Delta x_3 > \Delta^2$. Since $\Delta
x_3$ is bounded by the size of the box $\Sigma$, we find that $E$ must
be bounded above by ${\Sigma / \Delta^2}$ in order to keep the
excitations localized.  Combining these observations, we learn that
the physics of the NCSYM is natural in the range of energies
\beq {1 \over \Sigma} < E < {\Sigma \over \Delta^2} = {s \over
\Sigma}. \label{effrange}\eeq
It is important to emphasize that at the fundamental level the NCSYM
description does not break down above that scale. It is just becoming
very complicated because the relevant excitations are not localized
relative to the size of the torus.

Let us now consider the OSYM. Since this theory is an ordinary
conformal field theory, one can expect this description to remain
valid at arbitrarily high energies.  However, on the infra red side,
the momentum modes have a gap of the order of $1/\tilde{\Sigma}$.
Therefore, the range of naturalness of the OSYM description is bounded
in the infra red by $E = 1/\tilde{\Sigma} = s/\Sigma$.  Below this
energy, we have a two dimensional OSYM in the absence of the magnetic
flux. In the presence of the magnetic charge, however, there are
excitations which are much lighter than the gap of the momentum modes,
$1/\tilde{\Sigma}$.  These are the electric fluxes%
\footnote{ At an intuitive level, the electric fluxes behave
           differently in the magnetic background because they also
           carry momentum due to the Pointing vector $\vec{P}\sim
           \vec{B}\times\vec{E}$. See the appendix and
           \cite{Konechny:1998wv,vanBaal:1984ar} for a more detailed
           discussion.}
whose energies are quantized in the units of $1/\Sigma$.  Since the
Compton wave lengths of the states with such energies are much larger
than the size of the torus, these excitations are highly non-local and
hence it is very hard to describe their dynamics in terms of the OSYM.
In terms of the dual NCSYM description, these excitations are nothing
but the momentum modes (see appendix and
\cite{Konechny:1998wv,Pioline:1999xg}) whose dynamics is much simpler.

It should be emphasized that the physics of the cross-over on the
infra red side is different between the twisted SYM and the untwisted
SYM. In the untwisted SYM, the cross-over in the infra red is due to
the freezing of all but the zero momentum modes, whereas in the
twisted SYM, there are additional degrees of freedom due to the
electric fluxes. The dynamics of these degrees of freedom is given not
in terms of a dimensionally reduced theory, but by the 
4D NCSYM.

It should also be emphasized that the twisted SYM, the NCSYM, and the
dimensionally reduced SYM descriptions are natural in their respective
non-overlapping range of energies, but they combine to cover the full
range of energies.

\subsection{Large 't Hooft coupling and supergravity}

In the previous subsection, we assumed that the theory is weakly
coupled to argue that the duality between the NCSYM and the OSYM gives
rise to a hierarchy of phases.  We will now show that a similar
hierarchy appears also in the strongly coupled theory.  This suggests
that the structure of NCSYM-OSYM hierarchy is not sensitive to the value of
the coupling constant.
 
The supergravity solution which is dual to the NCSYM with large 't
Hooft coupling is \cite{Hashimoto:1999ut,Maldacena:1999mh}
\begin{eqnarray}\label{rt}
ds^2&=& \al \left\{
\frac{U^2}{\sqrt{\lambda}}(-dt^2+dx_1^2) +\frac{\sqrt{\lambda}
U^2}{\lambda + U^4 \Delta^4}(dx_2^2+dx_3^2)
+\frac{\sqrt{\lambda}}{U^2}dU^2 +\sqrt{\lambda}d \Omega_5^2 \right\},
\nonumber \\ 
e^{\phi}&=&\frac{\lambda}{4\pi N}
\sqrt{{\lambda \over \lambda + \Delta^4 U^4}}, \qquad
B_{23} = -{\alpha' \Delta^2 U^4 \over \lambda + \Delta^4 U^4}, \label{sugra}
\end{eqnarray} 
with periodicities $x_2 \sim x_2 +2 \pi \Sigma$ and $x_3 \sim x_3 +2
\pi \Sigma$.  In the infinite volume limit ($\Sigma \rightarrow
\infty$), the solution can be trusted everywhere for large $N$ and
$\lambda$.  For finite $\Sigma$, the supergravity description breaks
down at the ultra violet since the size of the torus is shrinking.
This implies that a different T-dual description takes over.  The
transition occurs when the invariant size of the torus is of the order
of the string length, so that the momentum modes are no longer the
lightest modes.  This happens at
\beq U=\frac{\lambda^{1/4}\Sigma}{\Delta^2}.  \eeq
Similarly, the size of the torus becomes small when $U$ gets smaller
than $\lambda^{1/4}/ \Sigma$.  This implies, using the UV/IR relation
\cite{Susskind:1998dq}, that the range of energies for which the
supergravity description of the NCSYM at large 't Hooft is valid is
\beq {1 \over \lambda^{1/4}} {1 \over \Sigma} < E <
\frac{1}{\lambda^{1/4}}{\Sigma \over \Delta^2}.  \eeq
This has basically the same form as (\ref{effrange}) up to the factor
of $\lambda^{1/4}$ which is a strong coupling effect.

Outside this range of energies, some other T-dual description is more
natural.  In the infra red the proper description is via the near
horizon limit of D1-branes \cite{Itzhaki:1998dd}.  This is just the
supergravity realization of the fact that the theory flows in the
infra red to a theory in two dimensions. This transition, which
involves a Gregory-Laflamme transition
\cite{Gregory:1993vy,Gregory:1994bj} in addition to the T-duality, was
discussed in \cite{Barbon:1998cr,Li:1998jy,Martinec:1998ja} and will not elaborated
on further here.

The more interesting transition for our discussion is the one in the
ultra violet.  The way T-duality acts on the supergravity solution is
the following \cite{Giveon:1994fu}. Define
\beq \rho = {\Sigma^2 \over \alpha'} \left(B_{23} + i \sqrt{G_{22}
 G_{33}}\right), \eeq
and act by an element of $SL(2,Z)$ according to
\beq \tilde \rho = {a \rho + b \over c \rho + d}. \eeq 
The dual $B$-field and the dual volume can be extracted from the real
and the imaginary parts of $\tilde \rho$, respectively.
\begin{eqnarray}
b_{23} & =& {\tilde{\Sigma}^2 \over \alpha'}\tilde{B}_{23} = \im
(\tilde \rho \nonumber) = {b d \lambda + (b \Delta^2 + a \Sigma^2) (d
\Delta^2 + c \Sigma^2) U^4
\over
d^2 \lambda + (d \Delta^2 + c \Sigma^2)^2 U^4},
\\
v_{23} &=& {\tilde{V}_{23} \over \alpha'}  =  \re( \tilde{\rho}) =
{\sqrt{\lambda} \Sigma^2 U^2
\over
d^2 \lambda + (d \Delta^2 + c \Sigma^2)^2 U^4} \label{rho}.
\end{eqnarray}
It is clear from (\ref{rho}) that generically, $v_{23}$ will always be
small for sufficiently large $U$. In the case where $\Theta =
\Delta^2/\Sigma^2 = 1/s$, however, the coefficient of the term in the
denominator of (\ref{rho}) proportional to $U^4$ can be set to zero by
choosing $c = -1$, $d=s$. This is the same element of $SL(2,Z)$ which
mapped the NCSYM to the OSYM in the previous subsection, and gives
rise to the following background
\begin{eqnarray}
ds^2&=& \al \left\{
\frac{U^2}{\sqrt{\lambda}}\left(-dt^2+dx_1^2+dx_2^2+dx_3^2
\rule{0ex}{3ex}\right) +\frac{\sqrt{\lambda}}{U^2}dU^2
+\sqrt{\lambda}d \Omega_5^2 \right\}, \nonumber \\
e^{\phi}&=&\frac{\lambda}{4\pi sN}, \qquad
B_{23} = -{1 \over s} {\alpha' \over \tilde{\Sigma}^2 },
\label{rationalfinal} 
\end{eqnarray} 
with periodicities $x_2 \sim x_2 +2\pi \tilde{\Sigma}$ and $x_3 \sim
x_3 + 2\pi \tilde{\Sigma}$.  The background geometry is exactly $AdS_5
\times S_5$ which implies \cite{Maldacena:1997re} that the theory is
OSYM in 4D.  The presence of the {\em constant} background $B$-field
gives rise to the twist in the corresponding boundary theory. In
particular, the constant
 $B$-field modifies the spectrum of closed string
winding modes and not of the momentum modes, just as the twist
modifies the spectrum of the electric fluxes and not of the local
momentum modes. The radius of $S^5$ remained the same as
(\ref{sugra}). On the other hand, the number of D3-branes is
multiplied by a factor of $s$, which can be seen either from the form
of the dilaton or the form of the RR field background which can be
found in \cite{Maldacena:1999mh}. The size of the torus, on the other
hand, is smaller by a factor of $s$.  All these features are in
agreement with the field theory results of equation (\ref{d2}) which
relate the NCSYM and the OSYM.

\begin{figure}
\centerline{\psfig{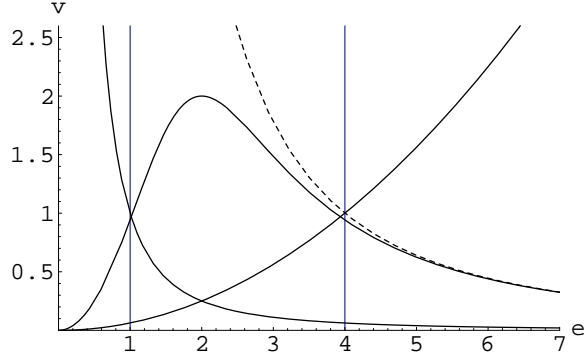}}
\caption{\label{figb} The different phases from the supergravity point
  of view.  Each of the curves in corresponds to the volume of the
  torus in the supergravity solution as a function of
  $e=U/\lambda^{1/4}$ where we absorbed the factor of $\lambda^{1/4}$
  into the definition of $e$ to account for the effect of strong
  coupling. We see that there are, like in the weakly coupled case,
  three ``phases,'' corresponding to the dimensionally reduced SYM,
  the NCSYM, and the twisted OSYM. This is to be compared against the
  case with $\tilde m=0$, which we illustrated in the same figure
  using the dotted line. Here, there are only two ``phases,''
  corresponding to the dimensionally reduced SYM and the OSYM. }
\end{figure}

\subsection{General $\Theta$}

In light of the analysis in the previous subsection for the case when
$\Theta = 1/s$, it is straight forward to understand what happens in
the case of a more general $\Theta$.  Consider first the case where
$\Theta = r/s$ is some rational number. Just as before, the NCSYM
description is expected to be the natural description in the range of
energies
\beq {1 \over \Sigma} < E < {\Sigma \over \Delta^2} = 
{s \over r} {1 \over \Sigma}, \eeq
and the dimensionally reduced description takes
over for $E < 1/\Sigma$. On the other hand, the OSYM description is
only valid at energies above $E = s / \Sigma$, and for $r > 1$, there
is a gap between $s/r\Sigma$ and $s/\Sigma$. Therefore, there must
be some other Morita equivalent description which is valid in this
intermediate regime.  Indeed, it is easy to see that for every energy
scale $E$, there is one and only one Morita equivalent description
which satisfies
\beq {1 \over \tilde{\Sigma}} < E < {\tilde{\Sigma} \over \tilde{\Delta}^2}.
\eeq
It is straight forward to search for the appropriate $SL(2,Z)$ dual
description using the relevant equations given in the appendix. For
example, for $\Theta = 13/47$ we find the following chain of natural
descriptions%
\footnote{In the following, we drop the dimensionally reduced SYM
	phase for brevity.}

\beq
\begin{array}{lcl}
\tilde{\Delta}^2 = {13 \over 47} \Sigma^2 &\qquad
{1 \over \Sigma} < E < {47 \over 13}{1 \over \Sigma}\qquad &
M=\left(\begin{footnotesize}\begin{array}{cc}1 & 0 \\ 0 & 1
\end{array}\end{footnotesize} \right) \\ \hline
\tilde{\Delta}^2 = {65 \over 2209} \Sigma^2 &
{47 \over 13}{1 \over \Sigma} < E < {47 \over 5}{1 \over \Sigma} &
M=\left(\begin{footnotesize}\begin{array}{cc}0 & 1 \\ -1 & 4
\end{array}\end{footnotesize}\right) \\ \hline
\tilde{\Delta}^2 = {10 \over 2209} \Sigma^2 &
{47 \over 5}{1 \over \Sigma} < E < {47 \over 2} {1\over  \Sigma} &
M=\left(\begin{footnotesize}\begin{array}{cc}-1 & 4 \\ -3 & 11 \end{array}\end{footnotesize}\right)  \\ \hline 
\tilde{\Delta}^2 = {2 \over 2209} \Sigma^2 &
{47 \over 2}{1 \over \Sigma} < E < {47 \over \Sigma} &
M=\left(\begin{footnotesize}\begin{array}{cc}-3 & 11 \\ -8 & 29 \end{array}\end{footnotesize}\right)  \\ \hline 
\tilde{\Delta}^2 = 0 &
{47 \over \Sigma} < E &
M=\left(\begin{footnotesize}\begin{array}{cc}-8 & 29 \\ -13 & 47 \end{array}\end{footnotesize}\right)  
\end{array} \label{th1347}
\eeq
where $M$ is an element of $SL(2,Z)$ defining the Morita dual relative
to the original theory with $\Theta = 13/47$.

Once the case of general rational $\Theta$ is understood, it is not
difficult to see what happens in the case of irrational
$\Theta$. Since for irrational $\Theta$ we can never reach the
$\tilde{\Theta} = 0$ theory by finite action of $SL(2,Z)$, the chain
of Morita equivalent theories will not terminate but continue
indefinitely. For example, for $\Theta = \gamma$ where $\gamma \approx
0.58$ is the Euler's constant, we find the following chain of natural
Morita equivalent descriptions.
\beq
\begin{array}{lcl}
\tilde{\Delta}^2 = 0.58 \Sigma^2 &
{1 \over \Sigma} < E < 1.7 {1 \over \Sigma} &
M=\left(\begin{footnotesize}
\begin{array}{cc}1 & 0 \\ 0 & 1 \end{array} 
\end{footnotesize}\right)  \\ \hline 
\tilde{\Delta}^2 = 0.089 \Sigma^2 &
1.7 {1 \over \Sigma} < E < 6.5 {1 \over \Sigma} &
M=\left(\begin{footnotesize}
\begin{array}{cc}0 & 1 \\ -1 & 2 \end{array}
\end{footnotesize}\right)  \\ \hline 
\tilde{\Delta}^2 = 0.0063 \Sigma^2 &
6.5 {1 \over \Sigma} < E <24.7 {1\over  \Sigma} &
M=\left(\begin{footnotesize}
\begin{array}{cc}-1 & 2 \\ -4 & 7 \end{array}
\end{footnotesize}\right)  \\ \hline 
\tilde{\Delta}^2 = 0.0003 \Sigma^2 &
\qquad 24.7 {1 \over \Sigma} < E < 131.5 {1 \over \Sigma} \qquad&
M=\left(\begin{footnotesize}
\begin{array}{cc}-4 & 7 \\ -15 & 26 \end{array}
\end{footnotesize}\right)  \\ \hline 
\hfil \vdots &
      \vdots &
\hfil \vdots
\end{array}\label{theulergamma}
\eeq

Let us pause and comment about the general structure of these chains of
theories. First, the parameters of two adjacent members of this chain
should satisfy an inequality
\beq {1 \over \Sigma} < {\Sigma \over \Delta^2} = {1 \over \tilde{\Sigma}} < {\tilde{\Sigma} \over \tilde{\Delta}^2}. \eeq
From this, it follows immediately that
\beq \tilde{\Sigma} < \Sigma, \qquad \tilde{\Delta} < \Delta, \eeq
which states that both the non-commutativity and the volume decreases
monotonically along the chain. Second, since all of these theories are
equivalent to one another at the microscopic level, they should all
have the same thermodynamic properties. Therefore, the relation
between the entropy $S$ and the temperature $T$
\beq S \sim \tilde{N}_\Theta^2 \tilde{\Sigma}^2 R
T^3, \qquad \tilde{N}_{\Theta} = \tilde N - \tilde \Theta \tilde m, \eeq
must be invariant under the $SL(2,Z)$ duality group, which we verify
in the appendix. Here, $R$ denotes the compactification radius in the
$x_1$ direction, which we take to be much larger then $\Sigma$.

The same basic structure can be seen also in the supergravity
description of the same model with large 't Hooft coupling.
\begin{figure}
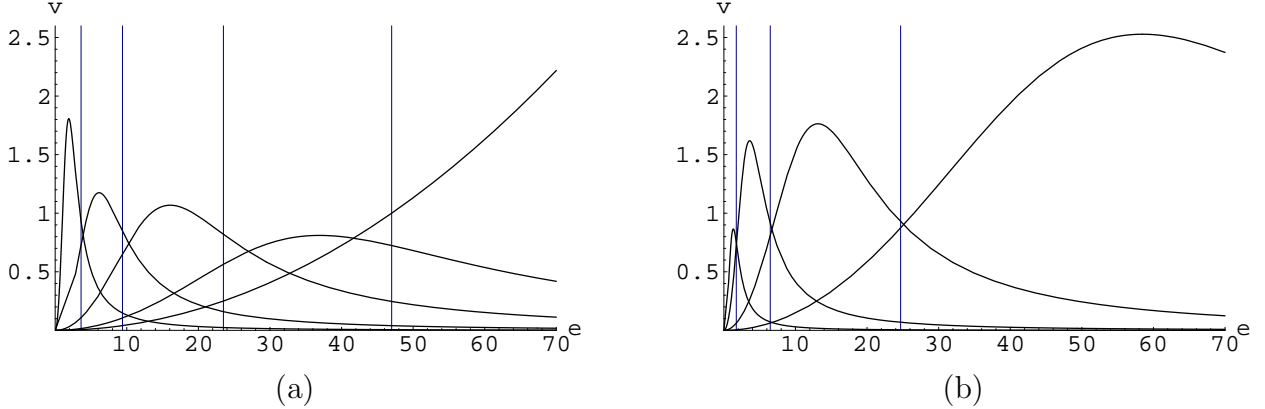

\parbox[b]{3in}{
\centerline{\psfig{file=th.13.47.epsi,width=3in}}
\centerline{(a)}
}\hfill
\parbox[b]{3in}{
\centerline{\psfig{file=th.eulergamma.epsi,width=3in}}
\centerline{(b)}}
\caption{Phases of Morita equivalent NCSYM for (a) $\Theta = 13/47$
and (b) $\Theta = \gamma \sim 0.58$. The left most peak illustrates
the dependence of the size of the torus as a function of $e$ of the
original NCSYM. Each successive peak corresponds to the size of the
torus of the Morita dual.\label{figa}}
\end{figure}
In figures \ref{figa}.a and \ref{figa}.b, we plot $v_{23}(e)$ for
the chain of natural Morita equivalent theories (\ref{th1347}) and
(\ref{theulergamma}) with $\Theta = 13/47$ and $\Theta = \gamma \approx
0.58$, respectively. The vertical lines delineate the range of
energies tabulated in (\ref{th1347}) and (\ref{theulergamma}).  The
left most peak in figures \ref{figa}.a and \ref{figa}.b illustrate the
dependence of the torus size as a function of $e$ of the supergravity
background corresponding to the NCSYM (\ref{sugra}).  Each successive
peaks correspond to the supergravity background of some Morita dual. A
given supergravity description is natural when $v(e)$ is of order one
or greater. Therefore, we find that the chain of natural Morita
equivalent theories (\ref{th1347}) and (\ref{theulergamma}) is also
natural from the point of view of supergravity. If $\Theta$ is
rational, eventually we will encounter $v(e)$ which grows like $e^2$
indefinitely (\ref{rationalfinal}).  This is the right most curve
illustrated in figure \ref{figa}.a and corresponds to the OSYM. For
irrational $\Theta$ like the one illustrated in figure \ref{figa}.b,
new peaks will appear indefinitely for arbitrarily large $e$.

\section{Conclusions}

The goal of the paper was to show that the theories which are related
by Morita equivalence have natural bounds in energies (\ref{effrange})
for which the dynamics of the effective degrees of freedom are most
local and therefore natural.  In this way, various Morita equivalent
theories form a hierarchy of phases, each being the appropriate
description for the physics at that scale.  We showed, for weakly as
well as for strongly coupled theories, that such a natural description
is unique at any given energies. For rational $\Th$ we find a finite
chain of natural descriptions which ends in the ultra violet by a
twisted OSYM. For irrational $\Th$ we find an infinite chain of
natural NCSYM descriptions.

We will close by commenting on an amusing application\footnote{Similar
ideas were considered in \cite{Bigatti:1998vf}.} of the NCSYM-OSYM
hierarchy to a version of large $N$ reduction of gauge theories
\cite{Eguchi:1982nm,Bhanot:1982cm,Parisi:1982gp,Gross:1982at}.  As we
described earlier,
\beq \left( \rule{0ex}{3ex} U(N)\ {\rm NCSYM}\ \Theta = 1/s \right) =
\left(\rule{0ex}{3ex}U(sN) \ {\rm OSYM}\ \tilde m = N \right)\eeq
at the microscopic level due to Morita equivalence. The theory on the
left hand side can be described to a good approximation by the usual
SYM (on the same space-time with the same gauge group and no twist)
 at energies
below the non-commutativity scale
\beq E < {1 \over \Delta} = {\sqrt{s} \over \Sigma}. \eeq
This follows from the fact that the $*$-product deforms the SYM by
irrelevant operators which do not affect the infra red dynamics.  If
we now take the limit $s \rightarrow \infty$, the theory on the left
hand side reduces to an ordinary $U(N)$ SYM on a torus of size
$\Sigma$. On the other hand, the theory on the right hand side becomes
a $U(sN)$ theory defined on $R^2 \times T^2$ with a twist $\tilde m =
N$.  The size of the $T^2$ goes to zero in the large $s$ limit, and in
this way, we have related $U(N)$ theory in four dimensions to a $U(s
N)$ theory in two dimensions. It is easy to generalize to a reduction
from four dimensions to zero dimensions by considering the case where
$x_0$ and $x_1$ are non-commutative as well.  It would be interesting
to see if such a relation derived using the Morita equivalence is
related to the traditional ideas of large $N$ reduction
\cite{Eguchi:1982nm,Bhanot:1982cm,Parisi:1982gp,Gross:1982at} as well
as to some more recent works \cite{Aoki:1999vr,Bars:1999av,Ambjorn:1999ts}.

\section*{Acknowledgments}

We would like to thank D. Gross, K. Hashimoto, B. Pioline, and
especially E. Silverstein for stimulating discussions at different
stages of the work. The work of AH is supported in part by the
National Science Foundation under Grant No. PHY94-07194. The work of
NI is supported in part by NSF grant No. PHY97-22022.


\section*{Appendix}

        {\setcounter{section}{1}
	 \gdef\thesection{\Alph{section}}}
        {\setcounter{equation}{0}}

In this appendix we will review some basic facts about  NCSYM and
the Morita equivalence. For concreteness, we will focus on ${\cal
N}=4$ $U(N)$ NCSYM on $R \times S_1 \times T^2$ with non-commutative
coordinates
\beq [x_\mu , x_\nu] =  i \theta_{\mu \nu}. \eeq
We will take $T^2$ to be a square torus with radii $\Sigma$, and take
the radius $R$ of the $S_1$ to be much larger than $\Sigma$. Such a
theory is defined by the action
\beq S = \tr \int d^4 x \, \left({1 \over 4 {g}_{YM}^2} ({F}_{\mu \nu}
+ \Phi_{\mu \nu}) * ({F}^{\mu \nu} + \Phi^{\mu \nu}) + \ldots
\right)\eeq
whose parameters, in addition to $\Delta$, $R$, and $\Sigma$, are the
rank of the gauge group $N$, the gauge coupling $g_{YM}^2$ and the
background fields $\Phi_{\mu \nu}$.  ``\ldots'' indicates the scalar
and the fermion terms. The non-commutativity parameter $\theta_{\mu
\nu}$ is encoded in the $*$-product
\beq f(x) * g(x) = \left. e^{ {i \theta_{\mu \nu} \over 2} {\partial
\over \partial x_\mu} {\partial \over \partial x'_\nu}} f(x) g(x')
\rule{0ex}{3ex}\right|_{x=x'}. \eeq
The $\Phi$ dependent term in the action is topological and its main
role is to twist \cite{'tHooft:1981sz} the vacuum. This parameter
appears in the construction of the NCSYM via the decoupling limit of
open strings on D3-branes as was explained in
\cite{Seiberg:1999vs}. The precise role played by $\Phi$ in mapping
T-duality to Morita equivalence was elucidated recently in
\cite{Pioline:1999xg}.

  We will
also restrict our attention to the case where only the 23 component
(along the $T^2$) of $\theta_{\mu \nu}$ and $\Phi_{\mu \nu}$
\beq \theta_{23} = 2 \pi  \Delta^2, \qquad \Phi_{23} = \Phi, \eeq
is non-vanishing, and introduce a dimensionless non-commutativity parameter
\beq \Theta = {\Delta^2 \over \Sigma^2}.\eeq

The spectrum of BPS states in this theory was studied extensively
\cite{Ho:1998hq,Morariu:1998qm,Hofman:1998pd,Hofman:1998iy,Hofman:1998iv,
Schwarz:1998qj,Konechny:1998wv,Konechny:1999rz,Konechny:1999tu,Pioline:1999xg}.
These states are labeled by the quantum numbers corresponding to the
momentum $p_i$, the electric flux $w_i$, and the magnetic flux
$m_{ij}$, and have the masses
\beq E = {\Sigma_1 \Sigma_2 \Sigma_3 \over \Sigma_i^2 \Sigma_j^2}
{(m_{ij} + N_\Theta \Phi_{ij})^2 \over 4 g_{YM}^2 N_\Theta} + {\Sigma_i^2
\over \Sigma_1 \Sigma_2 \Sigma_3} {g_{YM}^2 \over 2 N_\Theta} \left( w_i
+ \Theta_{ij} p_j \right)^2 + {1 \over N_\Theta} \sqrt{ {k_i^2 \over
\Sigma_i^2}} \label{bpsmass}, \eeq
where 
\beq N_\Theta = N + {1 \over 2 } m_{ij} \Theta_{ji}, \qquad
k_i = p_i N_\Theta - m_{ij}(w_j + \Theta_{jk} p_k), \label{lo2}\eeq
$\Sigma_i$ is the radius along the $x_i$ coordinate, and $i,j,k$ run
over $1, 2, 3$.  All of the terms in (\ref{bpsmass})
are positive definite. From the first term of (\ref{bpsmass}), we can
infer that the vacuum (lightest) topological sector is the one with
magnetic quantum numbers
\beq m_{12} = m_{13} = 0, \qquad m = m_{23} = \left\lceil -
{N \Phi \over 1-
\Phi \Theta} \right\rfloor \label{mphi},\eeq
where $\lceil x \rfloor$ denotes an integer whose value is closest to
$x$.  In this sector, (\ref{lo2}) simplifies to
\beq k_1 = N_\Theta p_1, \qquad k_i = p_i N - m \epsilon_{ij}
w_j. \label{lo}\eeq
The $k_i$'s label the spectrum of quadratic fluctuations around the
background \cite{Konechny:1998wv}. Note in particular that (\ref{lo})
is the defining relation of the Pointing vector in the vacuum ($k_i =
0$).

The energy of an excited  state  is given in the second and the third
terms of (\ref{bpsmass}). We will ignore the second term since it only
gives rise to a small perturbation when $g_{YM}$ is small to obtain 
from the third term
\beq E^2 = {1 \over R^2} p_1^2 + {1 \over N_\Theta^2 \Sigma^2} k_2^2 +
{1 \over N_\Theta^2 \Sigma^2} k_3^2.  \label{mass} \eeq
From (\ref{lo}) we see that $k_i$ takes on values which are integer
multiples of $\GCD (m,N)$. These states come with $(\GCD
(m,N))^2$-fold degeneracy \cite{vanBaal:1984ar} which is a remnant of
the $N^2$ degeneracy of the gauge particles in the $U(N)$ gauge
theory.\footnote{This is closely related to fact that the gauge group
of the untwisted NCSYM dual to this theory is $U(\GCD (m,N))$.}
Therefore, the entropy is
\beq S =\GCD (m,N)^2 \left( \frac{N_\Theta \Sigma}{\GCD (m,N)}
  \right)^2  R T^3 =
 N_\Theta^2 \Sigma^2 R T^3 .\label{entropy}\eeq

Recalling the construction of the NCSYM as a decoupling limit of
D3-branes in type IIB string theory wrapping $S_1 \times T^2$
\cite{Connes:1998cr,Douglas:1998fm,Seiberg:1999vs}, consider acting on
the $T^2$ by an element of the T-duality group $SL(2,Z) \times
SL(2,Z)$.  The full T-duality group survives the decoupling limit and
is identified as the group of Morita equivalences of the NCSYM. We
will concentrate on $SL(2,Z)$ subgroup which only acts on the K\"ahler
structure. An element of $SL(2,Z)$ will map a NCSYM with rank $N$,
twist $m$, coupling $g_{YM}$, period $\Sigma$, non-commutativity
$\Theta$, and background flux $\Phi$, to its Morita equivalent dual
with parameters $\tilde N$, $\tilde m$, $\tilde{g}_{YM}$, $\tilde
\Theta$, $\tilde \Phi$, and $\tilde \Sigma$ according to
\begin{eqnarray}\label{ol}
&&\tilde \Theta = {c + d \Theta \over a + b \Theta}, \qquad 
\tilde \Phi = (a + b \Theta)^2 \Phi - b ( a + b \Theta), \qquad 
\tilde{\Sigma} = (a + b \Theta) \Sigma, \nonumber \\
&&\tilde{g}_{YM}^2 = (a + b \Theta) g_{YM}^2, \qquad 
\left(\begin{array}{c} \tilde m \\ \tilde N \end{array}\right)
= \left(\begin{array}{cc} a & b \\ c & d \end{array}\right)
\left( \begin{array}{c} m \\ N \end{array}\right),
\end{eqnarray}
where $a$, $b$, $c$, and $d$ are integers defining the element of
$SL(2,Z)$. One can verify that the transformations of $m$ and $\Phi$
are consistent with (\ref{mphi}). The $SL(2,Z)$ invariance of the
entropy (\ref{entropy}) follows immediately from the transformation
 of $\Sigma$ and the fact that
$N_\Theta$ transforms according to
\beq \tilde{N}_\Theta = {N_\Theta \over (a + b \Theta)}, \eeq
which implies together with (\ref{ol}) that the entropy is indeed
invariant.  The momentum and the electric flux quantum numbers are
also exchanged under the $SL(2,Z)$
\beq \left(\begin{array}{c} \tilde p_i \\ \tilde w_i\end{array}\right)
= \left(\begin{array}{cc} a & b \\ c & d\end{array}\right)
\left(\begin{array}{c} p_i \\ w_i\end{array}\right).  \eeq

Note that a NCSYM with $\Theta = r/s$ for relatively prime integers
$r$ and $s$ and $\Phi=0$ is Morita equivalent to an OSYM with
\beq \tilde \Sigma = {1 \over s} \Sigma, \qquad \tilde N = s N, \qquad
\tilde{g}_{YM}^2 = {1 \over s} g_{YM}^2, \qquad \tilde \Phi = -{q
\over s}, \label{moritaeq}\eeq
by choosing $c = -r$, $d = s$, and taking $b=q$ to be the solution of
\beq q r = (1\ {\rm Mod}\ s) \label{mod}.\eeq
The solution of (\ref{mod}) is uniquely determined by requiring $q/s$
to take on values between zero and one.  For $r=1$, this is solved by
$q=1$ which is the case described extensively in sections 2.1 and
2.2. This is the precise statement of equivalence between the NCSYM
and the twisted OSYM when $\Theta$ is a rational number.

\begingroup\raggedright\endgroup

\end{document}